\documentclass[10pt]{article}

\usepackage[english]{babel}

\usepackage[T1]{fontenc}
\usepackage{newtxtext}

\usepackage[letterpaper,margin=1in]{geometry}

\usepackage{amsmath}
\usepackage{newtxmath}
\usepackage[version=4]{mhchem}
\usepackage{float}
\usepackage{graphicx}
\usepackage{subcaption}
\usepackage{xcolor}
\usepackage[noblocks]{authblk}
\usepackage[super,sort&compress,comma]{natbib}
\usepackage[colorlinks=true, allcolors=blue]{hyperref}

\DeclareUnicodeCharacter{2010}{-}
\DeclareUnicodeCharacter{2011}{\mbox{-}}

\renewenvironment{abstract}{%
  \begin{center}%
  \begin{minipage}{0.80\textwidth}%
  \small
  \setlength{\parindent}{0pt}%
  \setlength{\parskip}{0pt}%
}{%
  \par
  \end{minipage}%
  \end{center}%
  \vspace{0.8em}%
}

\DeclareRobustCommand{\affiliationblock}[1]{#1}
\title{\textbf{RS-CIDER: A non-local machine learning model for approximating screened hybrid functionals}}
\author[1,2,3]{Zhuotao Jin}
\author[3]{Mohamed S. Abdallah}
\author[3,5]{Boris Kozinsky}
\author[3,4]{Kyle Bystrom}
\affil[1]{\affiliationblock{Center for Computational Science and Engineering, Massachusetts Institute of Technology}}
\affil[2]{\affiliationblock{Department of Materials Science and Engineering, Massachusetts Institute of Technology}}
\affil[3]{\affiliationblock{John A. Paulson School of Engineering and Applied Sciences, Harvard University}}
\affil[4]{\affiliationblock{Initiative for Computational Catalysis, Flatiron Institute}}
\affil[5]{\affiliationblock{Robert Bosch LLC Research and Technology Center, Watertown}}

\date{}

\begin{document}
\maketitle

\begin{abstract}
Screened hybrid functionals such as HSE06 improve the description of band gaps, charge localization, and redox energetics relative to semilocal approximations, but their explicit Hartree--Fock exchange term is computationally costly for large, periodic systems, especially in plane-wave basis set calculations. Here we present RS-CIDER, a machine-learned non-local exchange functional that approximates the short-range Hartree--Fock exchange term in HSE06 by explicitly fitting both ground-state energies and single-particle energy levels. RS-CIDER combines scale-invariant semilocal and non-local density descriptors and can be evaluated self-consistently without explicitly applying the short-range Hartree--Fock exchange operator. RS-CIDER shows close agreement with HSE06 for molecular reaction energies and solid-state band gaps. Further tests across distinct materials show agreement between RS-CIDER and HSE06 for local magnetism, Cu--O phase competition, polaron localization, and neutral-defect energetics. For an Fe olivine, chemistry-specific fine-tuning recovers the HSE06 Li intercalation voltage. A timing benchmark shows that RS-CIDER reduces the measured per-SCF-step wall time by more than an order of magnitude relative to HSE06. Together, these molecular and solid-state results establish RS-CIDER as an efficient self-consistent machine-learned surrogate for HSE06.
\end{abstract}

\section{Introduction}

Semilocal exchange--correlation (xc) approximations underpin large-scale electronic-structure calculations, but their delocalization error can lead to errors in predicted band gaps, charge localization, and redox energetics~\cite{cohen_insights_2008,mori-sanchez_localization_2008}. Screened hybrid functionals mitigate many of these errors by incorporating a fraction of short-range Hartree--Fock exchange. For a given inverse-length-scale screening parameter $\omega$, the range-separation of the Coulomb operator is given by
\begin{equation}
    \frac{1}{r} = \frac{\text{erfc}(\omega r)}{r} + \frac{\text{erf}(\omega r)}{r}~\label{eq:range_sep},
\end{equation}
and the screened exchange is obtained by evaluating the exact exchange operator with the short-range part of the Coulomb interaction (the first term on the right-hand side of Eq~\ref{eq:range_sep}).
The most widely used screened hybrid is HSE06~\cite{heyd_hybrid_2003,krukau_influence_2006}, which mixes 25\% of the short-range exchange energy (at $\omega=0.11$ Bohr$^{-1}$) into the PBE functional~\cite{Perdew1996_3865}. In addition to its many other applications~\cite{Janesko2009,Henderson2011,Seo2015}, HSE06 has become the \emph{de facto} standard for charged point defect calculations in solids~\cite{Freysoldt2014,davidssonFirstPrinciplesPredictions2018a,lambertElectronicOpticalProperties2025}, largely due to its reasonable prediction of band gaps~\cite{borlido_large-scale_2019}. Despite substantial algorithmic advances~\cite{10.1021/acs.jctc.6b00092}, repeated evaluation of the non-local exchange operator remains a substantial computational cost in self-consistent hybrid-functional calculations, particularly for large or periodic systems.

Machine-learned exchange functionals offer a route to emulate non-local exact exchange at lower cost. For example, the CIDER framework for machine learning density functionals~\cite{10.1021/acs.jctc.1c00904,PhysRevB.110.075130} has been used to fit the unscreened, ``full-range'' exact exchange for both molecules and solids. It has also been extended to explicitly fit single-particle energies to improve the accuracy of ionization potentials, electron affinities, and band gaps~\cite{doi:10.1021/acs.jctc.4c00999}, but doing so required a more complex and computationally intensive set of input and was limited to molecular training data. Previous CIDER24X results suggested that the difficulty of fitting orbital energies partly reflects the limited representation of long-range exchange by its exponentially decaying SDMX descriptors~\cite{doi:10.1021/acs.jctc.4c00999}. By targeting short-range exchange, RS-CIDER need not represent this long-range contribution, providing a more tractable setting for incorporating occupation-response constraints with computationally efficient descriptors. However, at a fixed screening parameter \(\omega\), short-range exact exchange does not retain the simple homogeneous coordinate scaling of full-range exact exchange~\cite{levy1985}. This introduces an additional scale not represented by the original scale-invariant CIDER descriptors~\cite{10.1021/acs.jctc.1c00904,PhysRevB.110.075130} and motivates the local momentum-scale coordinate used in RS-CIDER.

Here we introduce RS-CIDER, which adapts the CIDER framework to approximate the short-range Hartree--Fock exchange term in HSE06. At the standard HSE06 mixing fraction $\alpha=0.25$, the learned short-range Hartree--Fock exchange is combined with the remaining PBE exchange and correlation terms according to the HSE06 mixing expression. To learn the behavior of screened exchange under uniform scaling, RS-CIDER augments the descriptor set from previous work~\cite{PhysRevB.110.075130} with a local inverse-length-scale descriptor derived from the density and kinetic energy density. The resulting functional can be evaluated self-consistently without the costly repeated calculation of the short-range Hartree--Fock exchange operator.

RS-CIDER shows close agreement with HSE06 for molecular reaction energies and solid-state band gaps, establishing fidelity to HSE06 in both molecular and periodic benchmarks. Distinct materials tests further probe the functional under self-consistent electronic-state selection and ionic relaxation, including CuO magnetism and Cu--O phase competition, electron and hole polarons, neutral defects in trigonal Se, and olivine intercalation voltages. A timing benchmark on perfect and vacancy-containing diamond cells demonstrates a substantial reduction in per-SCF-step wall time relative to HSE06. Taken together, these results demonstrate RS-CIDER as a promising self-consistent machine-learned surrogate for HSE06 for molecular and solid-state applications.

\section{Results}

\subsection{The RS-CIDER functional}

Earlier CIDER functionals learned unscreened, full-range exact exchange from semilocal and non-local descriptors \cite{PhysRevB.110.075130}. RS-CIDER instead targets $E_{\mathrm{x}}^{\mathrm{HF,SR}}(\omega)$, the short-range Hartree--Fock exchange defined using the standard HSE06 range-separation parameter $\omega=0.11~\mathrm{Bohr}^{-1}$, by learning a residual relative to PBE exchange,
\begin{equation}
\Delta E_{\mathrm{x}}^{\mathrm{SR}} = E_{\mathrm{x}}^{\mathrm{HF,SR}}(\omega)-E_{\mathrm{x}}^{\mathrm{PBE}},
\qquad
E_{\mathrm{x}}^{\mathrm{SRX\text{-}ML}} = E_{\mathrm{x}}^{\mathrm{PBE}}+\Delta E_{\mathrm{x}}^{\mathrm{SR,ML}}.
\label{eq:scider_construction}
\end{equation}
Here $\Delta E_{\mathrm{x}}^{\mathrm{SR,ML}}$ approximates the target residual. At the standard HSE06 mixing fraction $\alpha=0.25$, the learned short-range Hartree--Fock exchange is combined with the remaining PBE exchange and correlation terms according to the HSE06 mixing expression (Methods, Eq.~\eqref{eq:methods_deploy}); if the residual is reproduced exactly, the resulting functional is algebraically identical to HSE06.

\begin{figure}[!htbh]
\centering
\includegraphics[width=\linewidth]{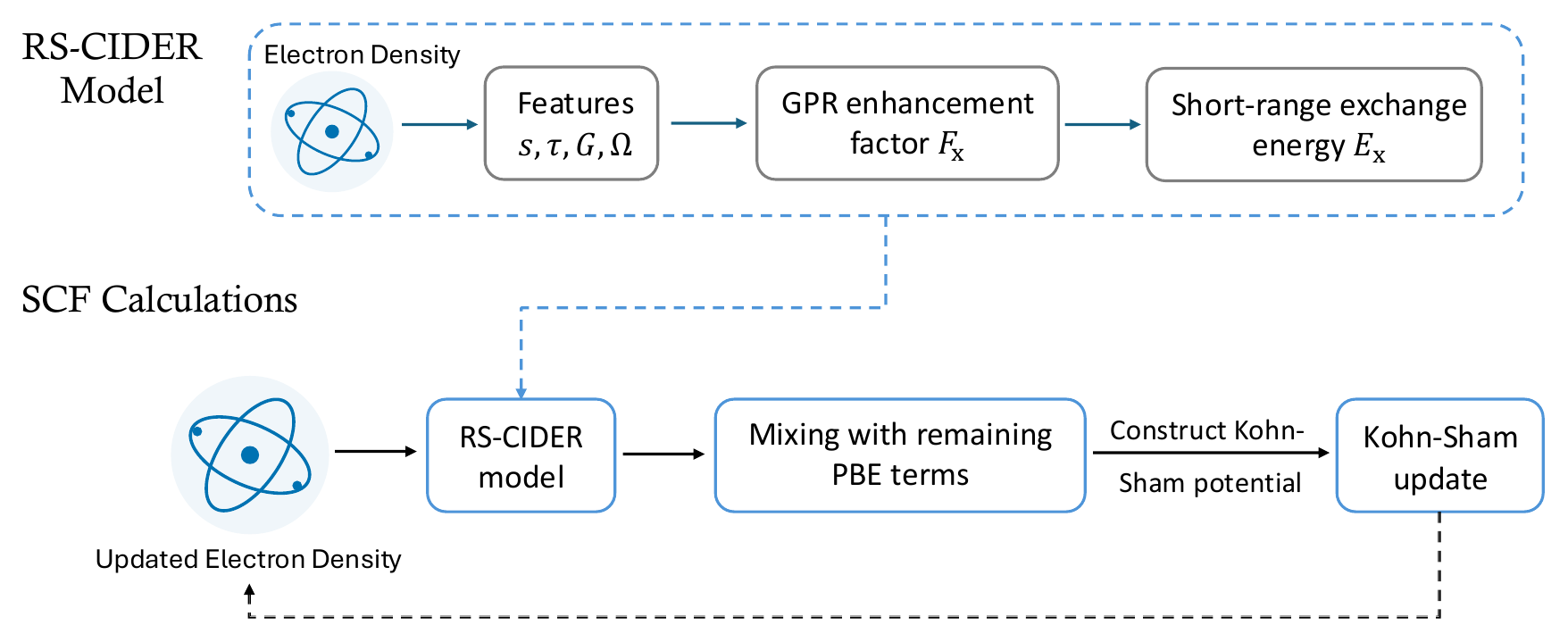}
\caption{RS-CIDER model and self-consistent evaluation. Density-derived features determine a Gaussian process regression (GPR) enhancement factor used to construct the learned short-range Hartree--Fock exchange. During self-consistent calculations, the trained exchange model is combined with the remaining PBE terms according to the HSE06 expression (Eq.~\eqref{eq:methods_deploy}), and the resulting exchange--correlation functional enters the Kohn--Sham update.}
\label{fig:concept}
\end{figure}

The representation combines the scale-invariant semilocal and non-local CIDER descriptors with a bounded coordinate derived from a local electronic momentum scale (see Eqs.~\ref{eq:methods_semilocal_coordinates}--\ref{eq:methods_momentum_coordinate}). RS-CIDER adapts the occupation-response training strategy used in earlier CIDER work to short-range exchange, combining short-range exchange energy residuals with occupation derivatives of the same contribution~\cite{doi:10.1021/acs.jctc.4c00999}. Figure~\ref{fig:concept} summarizes this learned exchange map and its use within the self-consistent Kohn--Sham cycle. Unless otherwise stated, all benchmarks below use the same general RS-CIDER model; chemistry-specific variants and calculations using a different exchange-mixing fraction are identified explicitly. Complete target definitions, training composition, and model hyperparameters are given in Methods and the Supplementary Information.

\subsection{Molecular reaction energies}

We first benchmarked RS-CIDER against HSE06 for molecular reaction energies using the GMTKN55~\cite{goerigk_look_2017} main-group chemistry benchmark database. Across the held-out test set of 1113 reactions in 52 subsets, RS-CIDER reproduced HSE06 reaction energies with a pooled mean absolute deviation (MAD) of 1.13 kcal\,mol$^{-1}$.

\begin{figure}[!htbh]
\centering
\includegraphics[width=\linewidth]{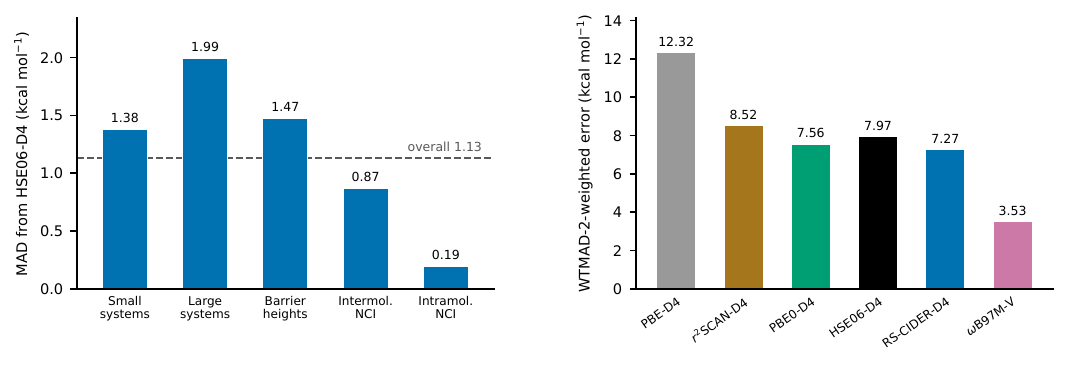}
\caption{Molecular reaction-energy benchmark on the held-out GMTKN55 partition comprising 1113 reactions from 52 subsets. Left: reaction-pooled MAD of RS-CIDER-D4 relative to HSE06-D4 within each of the five GMTKN55 categories; the dashed line is the MAD pooled over all held-out reactions. Right: Test-set WTMAD-2-type errors relative to the GMTKN55 reference energies for the displayed functionals (see Supplementary Information for the explicit definition).}
\label{fig:gmtkn55}
\end{figure}

Figure~\ref{fig:gmtkn55} separates fidelity to the HSE06 target from accuracy relative to the GMTKN55 reference energies. The category-pooled MADs relative to HSE06 range from 0.19 kcal\,mol$^{-1}$ for intramolecular noncovalent interactions to 1.99 kcal\,mol$^{-1}$ for large-system reaction and isomerization energies. The reference-energy comparison uses the established WTMAD-2 weighting introduced for GMTKN55, which accounts for differences in characteristic reaction-energy scales among its constituent subsets~\cite{goerigk_look_2017}; here, the same weighting is applied to the common held-out partition. All methods labeled ``-D4'' include DFT-D4 dispersion: RS-CIDER-D4 and HSE06-D4 use the HSE06 parameterization, whereas PBE-D4, r$^2$SCAN-D4, and PBE0-D4 use the parameterization of their parent functional. $\omega$B97M-V instead includes VV10 non-local correlation. With this weighting on the common held-out test partition, RS-CIDER-D4 gives 7.27~kcal\,mol$^{-1}$, comparable to PBE0-D4 (7.56~kcal\,mol$^{-1}$) and HSE06-D4 (7.97~kcal\,mol$^{-1}$), while outperforming PBE-D4 (12.32~kcal\,mol$^{-1}$) and r$^2$SCAN-D4 (8.52~kcal\,mol$^{-1}$). Thus, close agreement with HSE06 is accompanied by competitive accuracy against the molecular reference energies.

The five categories span small-system thermochemistry, large-system reactions and isomerizations, barrier heights, and both inter- and intramolecular noncovalent interactions. The RS-CIDER--HSE06 MAD remains below 2~kcal\,mol$^{-1}$ in every category and below 1~kcal\,mol$^{-1}$ for both noncovalent categories. The agreement therefore extends across distinct reaction-energy scales and interaction classes. Viewed together, the two panels establish complementary properties of the same general model: fidelity to HSE06 across chemically diverse reactions and competitive accuracy against the high-level reference energies.

\subsection{Solid-state band gaps}

We next evaluated solid-state band gaps with the same general RS-CIDER model for gapped solids in SOL62, with no solids band-gap data included in model training. RS-CIDER reproduced HSE06 with a mean absolute deviation of 0.20~eV. Against experiment, RS-CIDER approached HSE06-level agreement, with a mean absolute deviation of 0.92~eV compared with 0.76~eV for HSE06, while substantially outperforming PBE (1.93~eV) and r$^2$SCAN (1.42~eV).

The benchmark spans covalent semiconductors, ionic insulators, and rare-gas Ar, with HSE06 gaps ranging from below 1~eV to above 11~eV. Across this range, the mean signed RS-CIDER--HSE06 difference is $-0.10$~eV, alongside the 0.20~eV mean absolute deviation. These two measures show both limited overall bias and close material-by-material tracking over a broad range of gap magnitudes.

\begin{figure}[!htbh]
\centering
\includegraphics[width=\linewidth]{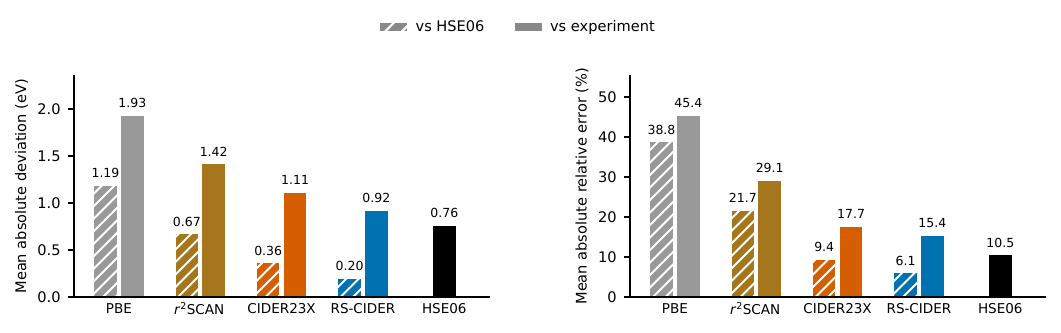}
\caption{Mean absolute deviations (left, eV) and mean absolute relative errors (right, \%) of calculated band gaps for gapped solids in SOL62 relative to HSE06 (hatched) and experiment (solid)~\cite{borlido_large-scale_2019}. The identically zero HSE06-to-HSE06 deviation is omitted. Per-material values and references are given in Supplementary Table~\ref{tab:solid_gaps}. CIDER23X refers to the PBE0/NL-MGGA-DTR model from previous work~\cite{PhysRevB.110.075130}.}
\label{fig:gaps}
\end{figure}

RS-CIDER followed HSE06's band gap underestimation for the widest-gap systems (CaO, NaCl, LiF, and Ar). The primary comparison is therefore the deviation from the HSE06 target, while experiment provides context for the accuracy of that reference.

\subsection{Correlated oxide: CuO}

Cupric oxide (\ce{CuO}) is an antiferromagnetic insulator driven by super-exchange effects~\cite{okeeffeMagneticSusceptibilityCupric1962,brownAntiferromagnetismCuOStudied1991}. However, PBE collapses to a closed-shell, non-magnetic ground-state. RS-CIDER and HSE06 both correctly identify the antiferromagnetic ground state. The resulting Cu moment of 0.63~$\mu_B$ is close to the HSE06 value of 0.69~$\mu_B$ and the experimental ordered moment of approximately 0.68~$\mu_B$~\cite{PhysRevB.38.174}.

\begin{figure}[!htbh]
\centering
\begin{subfigure}[t]{0.49\textwidth}
\centering
\includegraphics[width=\textwidth]{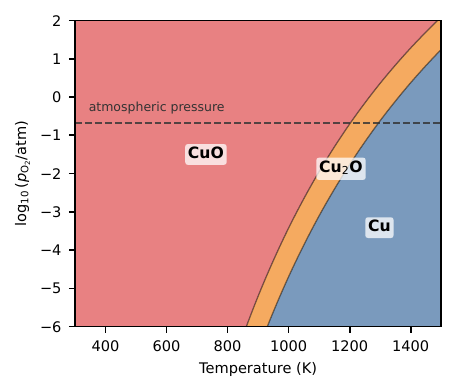}
\caption{HSE06}
\label{fig:cuo_hse06}
\end{subfigure}
\hfill
\begin{subfigure}[t]{0.49\textwidth}
\centering
\includegraphics[width=\textwidth]{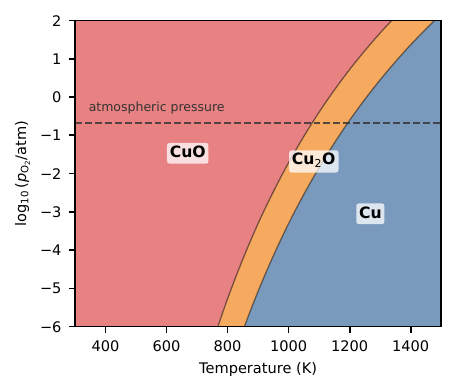}
\caption{RS-CIDER}
\label{fig:cuo_scider}
\end{subfigure}
\caption{Cu--O grand-potential phase diagrams. Filled regions identify the lowest-grand-potential phase among Cu, Cu$_2$O, and CuO at each temperature and oxygen partial pressure; the green dashed line marks the oxygen partial pressure of air. Energetic references and computational settings are given in the Supplementary Information.}
\label{fig:cuo_phase_diagram}
\end{figure}

We then constructed the Cu--O grand-potential phase diagram (Fig.~\ref{fig:cuo_phase_diagram}). RS-CIDER reproduced the HSE06 phase sequence among Cu, Cu$_2$O, and CuO, including the narrow intermediate Cu$_2$O stability region. Across most of the plotted pressure range, this stability window is shifted by roughly 100~K toward lower temperature relative to HSE06.

\subsection{Polaron localization}

We next turned to a distinct charge localization problem: small polarons, for which a localized carrier couples to an accompanying lattice distortion. We examined two complementary cases: an oxygen-vacancy electron polaron in rutile TiO$_2$ and a hole polaron in MgO.

\begin{figure}[!htbh]
\centering
\begin{subfigure}[t]{0.56\linewidth}
    \centering
    \includegraphics[width=\linewidth]{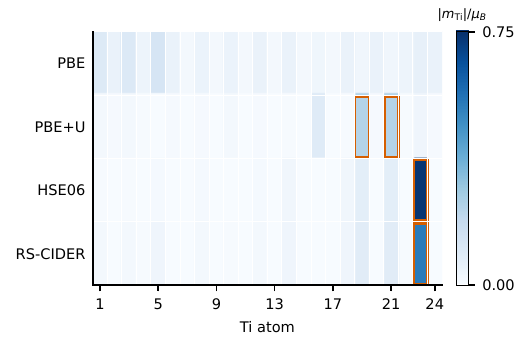}
    \caption{}
    \label{fig:polaron_tio2}
\end{subfigure}\hfill
\begin{subfigure}[t]{0.415\linewidth}
    \centering
    \includegraphics[width=\linewidth]{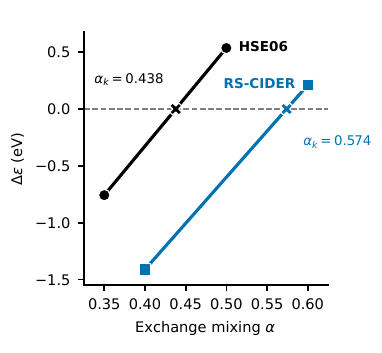}
    \caption{}
    \label{fig:polaron_mgo}
\end{subfigure}
\caption{Polaron tests in rutile TiO$_2$ and MgO. (a) Absolute local Ti magnetic moments in the $q=+1$ oxygen-vacancy cell, with rows denoting methods and columns following a common Ti-atom index. Orange outlines mark the two apical sites in the PBE+U row and the common HSE06/RS-CIDER site; the outline color does not encode a method. (b) Corrected polaron-level difference $\Delta\varepsilon$ versus exchange mixing $\alpha$ for the MgO hole polaron on the common PBE-relaxed lattice. Symbols denote the endpoint calculations, solid lines their two-point fits, and crosses the fitted zero crossings $\alpha_k$. Intermediate scan points and an additional RS-CIDER calculation at the fitted $\alpha_k$ are reported in the Supplementary Information.}
\label{fig:polaron}
\end{figure}

At the standard mixing fraction $\alpha=0.25$, RS-CIDER and HSE06 both localized the excess electron in the $q=+1$ TiO$_2$ oxygen-vacancy cell on the same Ti site (left panel of Fig.~\ref{fig:polaron}). The localized-site moments were 0.53 and 0.74~$\mu_B$, respectively, whereas the excess charge in the PBE solution remained diffuse. Starting from the same structure in the same fixed cell, the independently relaxed HSE06 and RS-CIDER geometries differed by only 0.018~\AA{} in the root-mean-square displacement over all 71 atoms. 

For MgO, we determined the exchange mixing fraction for each functional using the piecewise-linearity tuning protocol of Falletta and Pasquarello~\cite{falletta_polarons_2022}. Specifically, $\alpha$ was varied until the image-charge-corrected eigenvalue of the localized polaron level agreed between the charged and neutral systems, the endpoint condition associated with a piecewise-linear total energy through Janak's theorem~\cite{janak_proof_1978,PhysRevB.102.041115}. On the common PBE-relaxed lattice, this procedure gave $\alpha_k=0.438$ for HSE06 and $0.574$ for RS-CIDER, with both functionals localizing the hole (right panel of Fig.~\ref{fig:polaron}). After separately relaxing the MgO cell for each functional and refitting $\alpha_k$ on the resulting fixed cell, the RS-CIDER $\Delta$SCF formation energy closely reproduced HSE06 ($-0.34$ versus $-0.38$~eV). The complementary eigenvalue-based estimator showed a larger method dependence; the full values and correction details are reported in the ``Polaron localization''
section of Supplementary Information.

Thus, across an electron and a hole polaron in distinct oxides, RS-CIDER followed HSE06 in selecting localized solutions and closely matched the MgO $\Delta$SCF energetics.

\subsection{Neutral native defects in trigonal Se}\leavevmode\par

Neutral Se vacancies and interstitials probe the description of local coordination changes and structural relaxation in a solid. We evaluated the neutral Se vacancy ($V_{\mathrm{Se}}^{0}$) and Se interstitial ($\mathrm{Int}_{\mathrm{Se}}^{0}$), using the published HSE06+SOC calculations of Kavanagh \emph{et al.}~\cite{10.1039/d4ee04647a} as reference. In the own-bulk protocol, where the defect ions are relaxed at each functional's independently optimized bulk lattice, RS-CIDER gives formation energies of 1.47 and 0.95~eV for the vacancy and interstitial, respectively, compared with 1.53 and 0.82~eV from HSE06+SOC. RS-CIDER therefore closely reproduces both HSE06+SOC formation energies.

\begin{figure}[!htbp]
\centering
\includegraphics[width=0.88\linewidth]{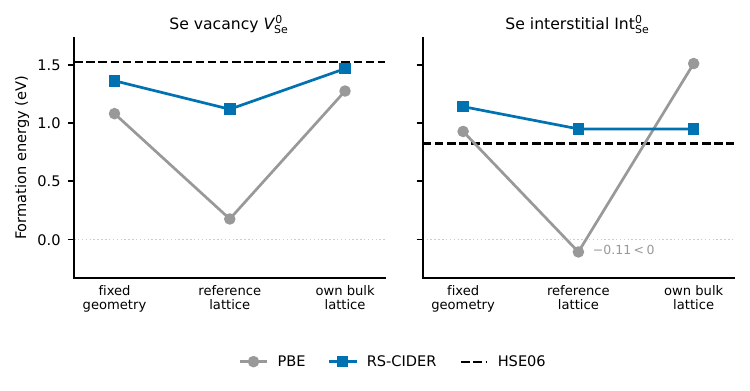}
\caption{Formation energies of the neutral Se vacancy (left) and interstitial (right) under three geometry protocols: all atoms fixed at the HSE06 reference geometry (``fixed geometry''), ions relaxed in the fixed reference cell (``reference lattice''), and ions relaxed at each functional's independently optimized bulk lattice (``own bulk lattice''). Lines connect the discrete protocols as visual guides. Dashed horizontal lines show the published HSE06+SOC reference values from Kavanagh \emph{et al.}~\cite{10.1039/d4ee04647a}.}
\label{fig:vacancy}
\end{figure}

To separate electronic-energy differences from structural response, we compared the three geometry protocols summarized in Fig.~\ref{fig:vacancy}. Across the relaxed protocols, RS-CIDER remains in good overall agreement with HSE06+SOC, although the reference-lattice vacancy shows a larger deviation. Because both defects are neutral, these comparisons require neither charged-supercell corrections nor cross-code band-edge alignment. The same D3 parameters were used throughout~\cite{grimme_consistent_2010}, and adding spin--orbit coupling to HSE06 changes the vacancy and interstitial formation energies by only 0.019 and less than 0.001~eV, respectively.

\subsection{Li intercalation voltages in olivine cathodes}

\begin{figure}[!htbh]
\centering
\includegraphics[width=0.7\linewidth]{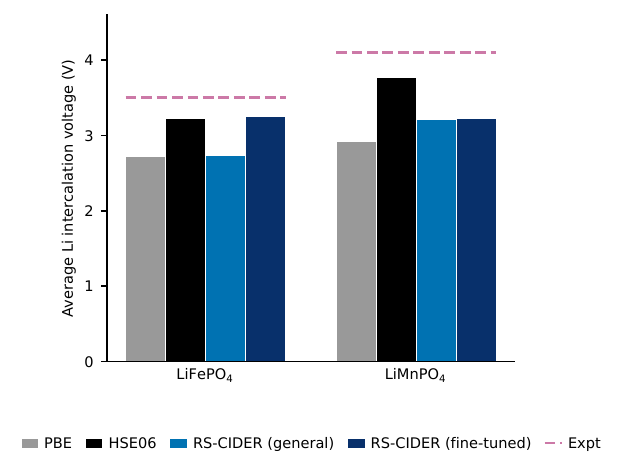}
\caption{Average Li intercalation voltages for LiFePO$_4$ and LiMnPO$_4$. Bars show PBE, HSE06, the general RS-CIDER model, and the corresponding chemistry-specific fine-tuned variant (RS-CIDER-Fe or RS-CIDER-Mn). Dashed horizontal lines mark the experimental voltages of 3.5 and 4.1~V, respectively~\cite{Padhi_1997,10.1021/cm030347b}.}
\label{fig:voltage}
\end{figure}

One of the key properties of Li-ion cathode materials is the intercalation voltage of Li, which is predicted inaccurately by semilocal DFT due to a combination of self-interaction and strong correlation error~\cite{PhysRevB.70.235121}. Here we examine the intercalation voltage of Li in the olivines LiFePO$_4$ and LiMnPO$_4$, which requires an accurate description of transition-metal redox chemistry. We evaluated intercalation voltage by referencing the total-energy difference between the lithiated and delithiated phases of the olivine to bulk Li (Fig.~\ref{fig:voltage}). For LiFePO$_4$, the general RS-CIDER model gives 2.74~V, significantly underestimating the value of 3.22~V from HSE06. To investigate this discrepancy, we trained the chemistry-specific RS-CIDER-Fe model by assigning greater statistical weight to short-range exchange energy and occupation-response constraints for the charge-state transitions Li$^{0}\!\rightarrow$Li$^{+}$, Fe$^{0}\!\rightarrow$Fe$^{+}$, and Fe$^{2+}\!\rightarrow$Fe$^{3+}$ in isolated atoms and ions. The resulting RS-CIDER-Fe model gives an intercalation voltage of 3.25~V, illustrating the potential for information learned from isolated species to transfer to solid-state redox energetics.

For LiMnPO$_4$, applying the analogous isolated-species fine-tuning changes the predicted voltage only from 3.22 to 3.23~V, compared with 3.78~V from HSE06. Bader charge analysis~\cite{HENKELMAN2006354} showed that the fine-tuned model and HSE06 gave similar net Mn-centered redox charge transfer values (0.372 and 0.346\,$e$, respectively). In the delithiated phase, the per-Mn charges from both the general and fine-tuned RS-CIDER models were within 0.05\,$e$ of HSE06. The voltage discrepancy therefore persists despite close agreement in the net Mn-centered redox charge transfer; we return to this contrast in the Discussion section.

\subsection{Computational cost}

\begin{figure}[!htbh]
\centering
\includegraphics[width=0.7\linewidth]{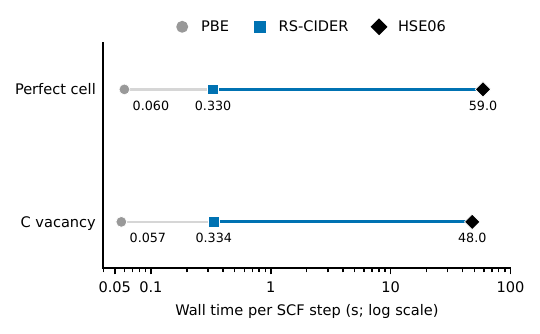}
\caption{Per-SCF-step wall times for the 8-atom conventional cubic diamond cell and the corresponding 7-atom C-vacancy cell, calculated with PBE, RS-CIDER, and HSE06 in GPAW using a $4\times4\times4$ $\Gamma$-centered Monkhorst--Pack $k$-point mesh and 32 MPI ranks. The horizontal axis is logarithmic. PBE and RS-CIDER points are medians of three paired calculations on the same physical node; HSE06 points are SCF-cycle averages from existing calculations with identical numerical settings on the same node type. Full timings and run metadata are reported in the Supplementary Information.}
\label{fig:cost}
\end{figure}

Finally, an illustrative timing benchmark on an 8-atom conventional cubic diamond cell and its 7-atom single-vacancy counterpart, with the same numerical settings and 32 MPI ranks used for each calculation, shows that RS-CIDER reduces the average wall time per completed SCF step by more than an order of magnitude relative to HSE06 (Fig.~\ref{fig:cost}). Plain PBE provides a semilocal cost reference in the figure.

\section{Discussion}

The same general RS-CIDER model shows close agreement with HSE06 for molecular reaction energies and solid-state band gaps. These complementary benchmarks span chemically diverse energy differences and self-consistent periodic electronic structure, showing that the CIDER density-functional construction can be extended from the full-range exact exchange treated previously~\cite{PhysRevB.110.075130} to the short-range Hartree--Fock exchange term in HSE06 while avoiding explicit exact exchange evaluation.

The materials applications extend the demonstrated agreement with HSE06 to electronic-state selection and structural response. RS-CIDER maintains an antiferromagnetic CuO solution and reproduces the HSE06 Cu--O phase sequence, selects the same localized carrier states as HSE06 in TiO$_2$ and MgO, and remains in good overall agreement with HSE06 for relaxed Se defects at both reference and method-specific lattices.

One area that requires further development is redox chemistry in battery materials. The RS-CIDER model did not initially match HSE06 for the intercalation voltages of Li in olivines. A modification of RS-CIDER fine-tuned to the ionization potentials and orbital energies of Fe and Li does recover the HSE06 Li intercalation voltage in \ce{LiFePO4}, illustrating that fine-tuning even on isolated species can potentially improve the description of redox in ionic materials. However, the analogous fine-tuning for Mn does not recover the HSE06 Li intercalation voltage in LiMnPO$_4$, indicating that further improvements are required beyond simple fine-tuning. The close agreement in Mn Bader charges indicates that the voltage discrepancy is not accompanied by a comparably large difference in net Mn-centered charge transfer. This suggests that net charge transfer alone is insufficient to recover the HSE06 voltage and points to aspects of the open-shell Mn exchange--correlation description not captured by this scalar charge measure. The Jahn--Teller-distorted Mn$^{3+}$ environment and its crystal-field-split $d$ manifold therefore provide an important regime for future model development~\cite{PRXEnergy.1.033003,10.1021/jp3122374}. More expressive descriptors could also potentially improve the accuracy of ML models for these systems~\cite{doi:10.1021/acs.jctc.4c00999}, though more work is required to develop efficient implementations of these features in plane-wave codes.

The practical significance of the agreement between RS-CIDER and HSE06 lies in replacing repeated evaluation of short-range Fock exchange with a learned density functional. In the timing benchmark on perfect and vacancy-containing diamond cells, RS-CIDER lowers the average wall time per completed SCF step by more than an order of magnitude relative to HSE06, with 32 MPI ranks used for each calculation. Taken together, these results demonstrate that RS-CIDER can serve as an efficient self-consistent machine-learned approximation to HSE06 for molecular and solid-state applications.

\section{Methods}

\subsection*{RS-CIDER functional}

RS-CIDER represents the short-range Hartree--Fock exchange as a PBE baseline plus a learned correction (Eq.~\eqref{eq:scider_construction}). The correction is a density functional constructed from a short-range LDA exchange-energy-density baseline and a dimensionless Gaussian-process enhancement factor,
\begin{equation}
\Delta E_{\mathrm{x}}^{\mathrm{SR,ML}}
= \int d^3r\; e_{\mathrm{x},\mathrm{SR}}^{\mathrm{LDA}}\!\big(\omega;n(\mathbf r)\big)
F_{\mathrm{x}}^{\mathrm{GPR}}\!\big(\mathbf x(\mathbf r)\big).
\label{eq:methods_residual_functional}
\end{equation}
RS-CIDER uses six bounded descriptor coordinates,
$\mathbf x=(x_s,x_\tau,x_{G_1},x_{G_2},x_{G_3},x_\Omega)$. Two semilocal
coordinates encode reduced-gradient and kinetic-energy-density information,
\begin{equation}
\begin{aligned}
s &= \frac{|\nabla n|}{2(3\pi^2)^{1/3}n^{4/3}},
&\qquad x_s &= \frac{\gamma_s s^2}{1+\gamma_s s^2},\\
\tau_0 &= \frac{3}{10}(3\pi^2)^{2/3}n^{5/3},
& x_\tau &= \frac{\tau-\tau_0}{\tau+\tau_0}.
\end{aligned}
\label{eq:methods_semilocal_coordinates}
\end{equation}
Here $\tau=\tfrac12\sum_p f_p|\nabla\phi_p|^2$ is the kinetic energy density, where $\phi_p$ and $f_p$ are Kohn--Sham orbitals and occupations. Three non-local coordinates are obtained from Gaussian convolutions of the density,
\begin{equation}
\begin{aligned}
&G_i(\mathbf r) = \int d^3r'\,n(\mathbf r')
\exp\!\left[-\big(a_\theta(\mathbf r')+a_i(\mathbf r)\big)
|\mathbf r-\mathbf r'|^2\right],
\qquad i=1,2,3,\\
&a_\nu(\mathbf r) = B_\nu n(\mathbf r)^{2/3}
+C_\nu\frac{\tau(\mathbf r)}{n(\mathbf r)},
\qquad \nu\in\{\theta,1,2,3\},\\
&\widehat G_i = \mathcal N_iG_i,\\
&x_{G_i} = -\frac12+\frac{\widehat G_i/2}{1+\widehat G_i/2}.
\end{aligned}
\label{eq:methods_nonlocal_coordinates}
\end{equation}
The normalization factors $\mathcal{N}_i$ are chosen so that $\widehat G_i = 2$ for the uniform electron gas (see Supplementary Information). The sixth coordinate is a bounded transformation of a local electronic momentum scale,
\begin{equation}
\Omega=n^{1/3}\sqrt{\frac12+\frac12\frac{\tau}{\tau_0}},
\qquad
x_\Omega=\frac{c_\Omega\Omega}{1+c_\Omega\Omega}.
\label{eq:methods_momentum_coordinate}
\end{equation}
The first five coordinates are invariant under uniform coordinate scaling. Here $\Omega$
combines the local density scale $n^{1/3}$ with the normalized kinetic-energy factor
$\sqrt{(1+\tau/\tau_0)/2}$. It is a local electronic inverse-length scale with the same
physical dimension as the range-separation parameter $\omega$, and $x_\Omega$ is its bounded,
monotonic transformation. Numerical mapping constants and Gaussian-exponent parameters are
given in the Supplementary Information, together with the exchange-spin-scaling convention
used for spin-polarized densities. Using the standard HSE06 parameters,
$\omega = 0.11$~bohr$^{-1}$ and $\alpha = 0.25$, the HSE06 and RS-CIDER
exchange--correlation energies can be written in parallel as
\begin{equation}
\begin{aligned}
&E_{\mathrm{xc}}^{\mathrm{HSE06}}
= (1-\alpha)\,E_{\mathrm{x},\mathrm{SR}}^{\mathrm{PBE}}(\omega)
+ E_{\mathrm{x},\mathrm{LR}}^{\mathrm{PBE}}(\omega)
+ E_{\mathrm{c}}^{\mathrm{PBE}}
+ \alpha\,E_{\mathrm{x}}^{\mathrm{HF,SR}}(\omega),\\
&E_{\mathrm{xc}}^{\text{RS-CIDER}}
= (1-\alpha)\,E_{\mathrm{x},\mathrm{SR}}^{\mathrm{PBE}}(\omega)
+ E_{\mathrm{x},\mathrm{LR}}^{\mathrm{PBE}}(\omega)
+ E_{\mathrm{c}}^{\mathrm{PBE}}
+ \alpha\,E_{\mathrm{x}}^{\mathrm{SRX\text{-}ML}}.
\end{aligned}
\label{eq:methods_deploy}
\end{equation}
Here $E_{\mathrm{x},\mathrm{SR}}^{\mathrm{PBE}}(\omega)$ and $E_{\mathrm{x},\mathrm{LR}}^{\mathrm{PBE}}(\omega)$ are the short- and long-range components of PBE exchange under the same $\mathrm{erfc}$/$\mathrm{erf}$ partition of the Coulomb kernel (Eq.~\eqref{eq:range_sep}). RS-CIDER retains the HSE06 mixing expression and replaces the short-range Hartree--Fock exchange with the learned exchange defined in Eq.~\eqref{eq:scider_construction}. The two expressions coincide when $E_{\mathrm{x}}^{\mathrm{SRX\text{-}ML}}=E_{\mathrm{x}}^{\mathrm{HF,SR}}(\omega)$, and setting $\alpha = 0$ recovers PBE. The GPR derivation appears in the Supplementary Information.

In the MgO polaron analysis, $\alpha$ is treated as tunable, playing the same role as the exact-exchange fraction of a hybrid functional. For the MgO hole polaron, we evaluated the formation energy using two complementary expressions. The $\Delta$SCF expression uses the finite-size-corrected total-energy difference between the charged polaron and neutral bulk cells, with the removed electron referenced to the valence-band edge. The eigenvalue-based expression combines the finite-size-corrected total-energy difference between the neutral polaron and bulk cells with the corrected localized-polaron eigenvalue. The two expressions coincide in the piecewise-linear limit, and their difference
measures residual nonlinearity.

\subsection*{Training targets}

The part of RS-CIDER that is machine-learned is the short-range exchange residual $\Delta E_{\mathrm{x}}^{\mathrm{SR,ML}}$ of Eq.~\eqref{eq:scider_construction}. $\Delta E_{\mathrm{x}}^{\mathrm{SR,ML}}$ was trained to reproduce the exact $\Delta E_{\mathrm{x}}^{\mathrm{SR}}$ for linear combinations of short-range exchange energies and their derivatives with respect to selected frontier orbital occupations~\cite{doi:10.1021/acs.jctc.4c00999}. For system $j$, let $\Delta E_{\mathrm{x},j}^{\mathrm{SR}}=E_{\mathrm{x},j}^{\mathrm{HF,SR}}(\omega)-E_{\mathrm{x},j}^{\mathrm{PBE}}$. The two basic forms for a training target $r$ are
\begin{equation}
\begin{aligned}
&y_r^{(E)} = \sum_j c_{rj}\,\Delta E_{\mathrm{x},j}^{\mathrm{SR}},\\
&y_r^{(d)} = \sum_{j,p} d_{rjp}\,
      \frac{\partial \Delta E_{\mathrm{x},j}^{\mathrm{SR}}}{\partial f_{jp}},
\end{aligned}
\label{eq:methods_training_labels}
\end{equation}
where $c_{rj}$ and $d_{rjp}$ specify the stoichiometric and occupation-response combinations, respectively, and $f_{jp}$ is the occupation of a selected frontier orbital with index $p$ in system $j$. Exchange energy targets like reaction energies and barrier heights use the first equation $y_r^{(E)}$, while orbital energy targets use the second form $y_r^{(d)}$. We also include so-called ``piecewise quadraticity constraints'' introduced in previous work~\cite{doi:10.1021/acs.jctc.4c00999}, which use a combination of $y_r^{(E)}$ and $y_r^{(d)}$ to target the physically correct behavior of exchange for a fractional number of electrons. Reference values for the uniform electron gas are included as additional training data, but are not enforced as exact constraints in this work. The explicit occupation-response combinations are given in the Supplementary Information.

\subsection*{Training procedure}

The model was fit by Gaussian-process regression (GPR) in a low-rank representation, with control points selected from training-system descriptor samples by pivoted Cholesky decomposition. The training systems span molecular and atomic chemistry, 3d-transition-metal ionization and spin states, and a small set of bulk-solid cohesive data. The training components and numbers of supervised constraints for the general RS-CIDER model are summarized in Table~\ref{tab:trainsets}. Detailed dataset composition, target definitions, control-point selection, and noise settings are reported in the Supplementary Information.

\begin{table}[ht]
\centering
\footnotesize
\caption{Training components and numbers of supervised constraints used for RS-CIDER. The target column identifies the corresponding short-range exchange constraint. IP and EA denote ionization potential and electron affinity, respectively. The listed constraints exclude the separate uniform gas targets described in the text.}
\label{tab:trainsets}
\begin{tabular}{l r l}
\hline
Training component & \# & Target \\
\hline
3d-TM ionization stoichiometries       &  60 & Exchange energy difference \\
3d-TM piecewise quadraticity           &  60 & Residual combining energy and occupation derivatives \\
3d-TM generalized Koopmans             &  60 & Occupation derivative residual \\
G21 IP piecewise quadraticity          &  36 & Residual combining energy and occupation derivatives \\
G21 EA piecewise quadraticity          &  25 & Residual combining energy and occupation derivatives \\
G21 frontier generalized Koopmans      &  61 & Occupation derivative residual \\
G21 ionization stoichiometries         &  36 & Exchange energy difference \\
G21 electron-affinity stoichiometries  &  25 & Exchange energy difference \\
GMTKN55 reaction stoichiometries       & 225 & Exchange energy differences \\
W4-11 atomization stoichiometries      & 140 & Exchange energy differences \\
W4-11 frontier occupations             & 456 & Occupation derivatives \\
Closed/open-shell reference atoms      &  45 & Exchange energies \\
Bulk-solid cohesive stoichiometries    &   9 & Exchange energy differences \\
\hline
Total                           & 1238 & \\
\hline
\end{tabular}
\end{table}

\subsection*{Electronic-structure calculations}

Molecular benchmark calculations (GMTKN55) for RS-CIDER and HSE06 used PySCF~\cite{sun_recent_2020} with the def2-QZVPPD basis~\cite{rappoport_property-optimized_2010} and DFT-D4 dispersion~\cite{caldeweyher_generally_2019}; additional PySCF comparison data used the same basis with the dispersion treatment indicated by each functional label. Periodic RS-CIDER calculations were performed self-consistently with GPAW~\cite{Enkovaara2010} in a plane-wave PAW representation, with the CiderPress code~\cite{ciderpress} used to evaluate the CIDER functionals. For the solid-state band-gap benchmark, the PBE, r$^2$SCAN, earlier CIDER, and HSE06 values were taken from Bystrom and Kozinsky~\cite{PhysRevB.110.075130}. Reference calculations for the other periodic applications used GPAW or VASP~\cite{kresse_efficient_1996} as specified in the Supplementary Information. Unless otherwise stated, HSE06 used 25\% short-range exact exchange with the standard screening parameter $\omega = 0.11~\mathrm{bohr}^{-1}$. Per-system plane-wave cutoffs, k-point meshes, supercell sizes, magnetic configurations, and RS-CIDER-specific settings are listed in the Supplementary Information.

\subsection*{Intercalation voltage}

The average Li intercalation voltage was computed as
\begin{equation}
V = \frac{E(\text{delithiated}) + n\,\mu_{\mathrm{Li}} - E(\text{lithiated})}{n e},
\label{eq:methods_voltage}
\end{equation}
where $\mu_{\mathrm{Li}}$ is the bulk-lithium chemical potential, taken as the energy per atom of bcc lithium, $n$ is the number of Li removed, and $e$ is the elementary positive charge.

\section*{Acknowledgements}
We acknowledge support from Tata Steel Ltd, the National Science Foundation Office of Advanced Cyberinfrastructure (OAC) under Award No.~2118201, and the U.S. Department of the Navy, Office of Naval Research under Award No.~N00014-20-1-2418. The authors acknowledge computing resources provided by the Harvard University FAS Division of Science Research Computing Group. The authors thank Se\'an R. Kavanagh and Hao Tang for helpful discussions. The authors also thank Stefano Falletta for sharing a script for generating the charge-image corrections for the polaron calculations. The Flatiron Institute is a division of the Simons Foundation.

\clearpage
\appendix
\section{Supplementary Information}
\subsection*{RS-CIDER functional form}

RS-CIDER learns the short-range Hartree--Fock exchange (SRX) of the HSE06 hybrid functional,
defined by the error-function screening kernel
$\mathrm{erfc}(\omega|\mathbf r-\mathbf r'|)/|\mathbf r-\mathbf r'|$ with
$\omega = 0.11~\mathrm{bohr}^{-1}$, and represents it through
a delta-learning model relative to a PBE exchange baseline,
\begin{equation}
\Delta E_{\mathrm{x}}^{\mathrm{SR}} \equiv E_{\mathrm{x}}^{\mathrm{HF,SR}}(\omega) - E_{\mathrm{x}}^{\mathrm{PBE}},
\label{eq:si_delta}
\end{equation}
where $E_{\mathrm{x}}^{\mathrm{HF,SR}}(\omega)$ is the short-range Hartree--Fock exchange before
application of the HSE06 mixing fraction. The residual is constructed by multiplying the short-range LDA by a machine-learned enhancement factor,
\begin{equation}
\Delta E_{\mathrm{x}}^{\mathrm{SR,ML}} = \sum_g w_g\, e_{\mathrm{x},\mathrm{SR}}^{\mathrm{LDA}}(\omega;n_g)\,
F_{\mathrm{x}}^{\mathrm{GPR}}(\mathbf x_g),
\label{eq:si_mul_baseline}
\end{equation}
where $w_g$ is the integration weight at real-space grid point $g$, $\mathbf x_g$ is the
descriptor vector at $g$, and $F_{\mathrm{x}}^{\mathrm{GPR}}$ is the dimensionless enhancement factor
learned by a Gaussian-process regression. The machine-learned short-range exchange is
$E_{\mathrm{x}}^{\mathrm{SRX\text{-}ML}} = E_{\mathrm{x}}^{\mathrm{PBE}} + \Delta E_{\mathrm{x}}^{\mathrm{SR,ML}}$. The
RS-CIDER functional combines this learned exchange with PBE using $\mathrm{xmix}=0.25$
(Eq.~\eqref{eq:methods_deploy}). The PBE baseline is subtracted when the residual labels are constructed and added back to form $E_{\mathrm{x}}^{\mathrm{SRX\text{-}ML}}$; it is not part of the Gaussian-process covariance kernel.

The six bounded coordinates are defined in
Eqs.~\eqref{eq:methods_semilocal_coordinates}--\eqref{eq:methods_momentum_coordinate}.
For connection to standard meta-GGA notation, the iso-orbital indicator is
$\alpha=(\tau-\tau_W)/\tau_0$, with $\tau_W=|\nabla n|^2/(8n)$, so that
$\tau/\tau_0=\alpha+\tfrac53s^2$. The reduced-gradient coordinate uses
$\gamma_s=0.243$, which Becke chose empirically to fit the exact exchange energies of noble gas
atoms~\cite{Becke_1986}.

The first five coordinate maps follow the CIDER construction~\cite{PhysRevB.110.075130}.
For the Gaussian exponents in Eq.~\eqref{eq:methods_nonlocal_coordinates}, the  parameters are equivalently written
\begin{equation}
\begin{aligned}
a_\theta &= \left(\frac{\pi}{2^{5/3}}-\frac{C_F}{16}\right)n^{2/3}
+\frac1{16}\frac{\tau}{n},
\qquad C_F=\frac3{10}(3\pi^2)^{2/3},\\
a_i &= \lambda_i a_\theta,
\qquad (\lambda_1,\lambda_2,\lambda_3)=\left(\frac12,1,2\right).
\end{aligned}
\label{eq:si_nl_exponents}
\end{equation}
The uniform electron gas normalization constants in
Eq.~\eqref{eq:methods_nonlocal_coordinates}, chosen so that $\widehat G_i=2$ and
$x_{G_i}=0$ for the uniform gas, are
\begin{equation}
(\mathcal N_1,\mathcal N_2,\mathcal N_3)
=\left(\frac{3\sqrt3}{8},1,\frac{3\sqrt6}{4}\right).
\label{eq:si_nl_norms}
\end{equation}
The momentum-scale map in Eq.~\eqref{eq:methods_momentum_coordinate} uses
$c_\Omega=0.1$~bohr. All descriptor quantities are evaluated in atomic units. The fixed
range-separation parameter $\omega$ enters the short-range exchange target, the short-range
LDA baseline, and the PBE short-/long-range partition.
For spin-polarized densities, the six coordinates are
evaluated separately in each channel using the exchange-spin-scaled quantities
$n=2n_\sigma$, $\nabla n=2\nabla n_\sigma$, and $\tau=2\tau_\sigma$. The channel
contributions are then combined according to
$E_{\mathrm{x}}[n_\uparrow,n_\downarrow]=\tfrac12E_{\mathrm{x}}[2n_\uparrow]+\tfrac12E_{\mathrm{x}}[2n_\downarrow]$.

\subsection*{Gaussian-process training}

The enhancement factor $F_{\mathrm{x}}^{\mathrm{GPR}}$ is fit to residuals in short-range
exchange energies (Eq.~\ref{eq:si_mul_baseline}). At grid
point $g$ of system $i$, define the multiplicative baseline
$m_g^i=e_{\mathrm{x},\mathrm{SR}}^{\mathrm{LDA}}(\omega;n_g^i)$, so that the exchange energy label is
$E_{\mathrm{x}}^i=\sum_{g\in i}w_g^i m_g^i F_{\mathrm{x}}^{\mathrm{GPR}}(\mathbf x_g^i)$. Let
$\kappa(\mathbf x,\mathbf x')=\mathrm{Cov}[F_{\mathrm{x}}^{\mathrm{GPR}}(\mathbf x),
F_{\mathrm{x}}^{\mathrm{GPR}}(\mathbf x')]$ denote the covariance kernel of the dimensionless
enhancement factor. Bilinearity of covariance then gives, for any two training systems $i$
and $j$,
\begin{equation}
\mathrm{Cov}(E_{\mathrm{x}}^i, E_{\mathrm{x}}^j) = \sum_{g\in i}\sum_{h\in j}
w_g^i\,w_h^j\,m_g^i m_h^j\,
\kappa(\mathbf x_g^i,\mathbf x_h^j),
\label{eq:si_ecov}
\end{equation}
Orbital-response labels are derivatives with respect to orbital occupation number of the same exchange contribution and
therefore enter the GP as linear functionals through the chain rule. Janak's
theorem~\cite{janak_proof_1978} relates the corresponding total-energy occupation derivatives,
of which this exchange response is one contribution, to frontier-orbital eigenvalues. The
corresponding control-point projection differentiates the complete product
$m_g^i\kappa(\mathbf x_g^i,\tilde{\mathbf x}_a)$ with respect to occupation, including the
occupation derivatives of both the LDA baseline and the descriptor vector. Exchange energy and
occupation-response constraints are combined in a single covariance matrix for joint fitting.
A complete derivation of the orbital-response construction is given in the
CIDER24X work~\cite{doi:10.1021/acs.jctc.4c00999}.

The real-space grid for each system carries $\sim 10^5$ points, so forming the full covariance
(Eq.~\ref{eq:si_ecov}) for every pair of systems is intractable; it is replaced by a
Nystr\"om low-rank approximation built on a small set of control points
$\tilde{\mathbf x}_a$ in feature space,
\begin{equation}
\mathrm{Cov}(E_{\mathrm{x}}^i, E_{\mathrm{x}}^j) \approx K_{ij} \equiv
(\tilde{\mathbf k}^i)^{\!\top}\, \tilde{\mathbf K}^{-1}\, \tilde{\mathbf k}^j,
\label{eq:si_nystrom}
\end{equation}
with $(\tilde{\mathbf K})_{ab}=\kappa(\tilde{\mathbf x}_a,\tilde{\mathbf x}_b)$ and
$(\tilde{\mathbf k}^i)_a=\sum_{g\in i}w_g^i m_g^i
\kappa(\mathbf x_g^i,\tilde{\mathbf x}_a)$.
Candidate points are obtained by downsampling the real-space grids of all training systems.
After spin-channel expansion, control points are selected automatically from the resulting
pool by pivoted incomplete Cholesky decomposition on the
normalised kernel matrix, with stopping tolerance $\mathtt{ctrl\_tol} = 10^{-3}$. The
predictor in the control-point basis is
\begin{equation}
F_{\mathrm{x}}^{\mathrm{GPR}}(\mathbf x_*) =
\sum_a \kappa(\mathbf x_*,\tilde{\mathbf x}_a)\,\alpha_a, \qquad
\boldsymbol\alpha = \tilde{\mathbf K}^{-1}\sum_i \tilde{\mathbf k}^i\,
\big\{[\mathbf K + \Sigma_{\mathrm{noise}}]^{-1}\mathbf y\big\}_i,
\label{eq:si_predictor}
\end{equation}
where $\mathbf K$ is the Nystr\"om label Gram matrix of Eq.~\eqref{eq:si_nystrom},
$\Sigma_{\mathrm{noise}}$ the diagonal label-noise covariance matrix, and $\mathbf y$ the stacked label
vector; for an occupation-derivative label, $\tilde{\mathbf k}^i$ is replaced by the
occupation derivative of the complete projection
$\sum_g w_g^i m_g^i\kappa(\mathbf x_g^i,\tilde{\mathbf x}_a)$ described
above~\cite{doi:10.1021/acs.jctc.4c00999}.

We use an additive Gaussian enhancement-factor kernel $\kappa$ with squared-exponential
components that support an efficient low-dimensional spline representation for self-consistent
evaluation~\cite{PhysRevB.110.075130}; length scales $l_i$ are set to the standard deviations
of the corresponding features over the downsampled descriptor samples used to form the
control-point candidate pool.

The resulting learned residual is evaluated as a single real-space integral of the
dimensionless enhancement factor on the short-range LDA baseline,
\begin{equation}
\Delta E_{\mathrm{x}}^{\mathrm{SR,ML}}
= \int d^3 r\; e_{\mathrm{x},\mathrm{SR}}^{\mathrm{LDA}}\!\big(\omega; n(\mathbf r)\big)\,
F_{\mathrm{x}}^{\mathrm{GPR}}\!\big(\mathbf x(\mathbf r)\big),
\quad
F_{\mathrm{x}}^{\mathrm{GPR}}(\mathbf x_*) = \sum_a \alpha_a\, \kappa(\mathbf x_*, \tilde{\mathbf x}_a),
\label{eq:si_deploy}
\end{equation}
with the same control-point coefficients as Eq.~\eqref{eq:si_predictor}.

\subsection*{Training data}

The RS-CIDER functional was fit by Gaussian-process regression to 1238 supervised constraints defined by thirteen reaction sets over eight sets of systems (Table~\ref{tab:trainsets}). Descriptor samples for the GP representation were drawn from the same system sets, and separate uniform gas constraints were added independently. The system sets include 3d transition metal ions and their spin states, the G21IP and G21EA molecules, the W4-11 molecules, a GMTKN55 reaction pool included in the control point candidate pool,
reference atoms, and bulk solids with their constituent free atoms~\cite{goerigk_look_2017,jmlm235,PhysRevB.110.075130,doi:10.1021/acs.jctc.4c00999}.
Each supervised constraint is a reaction-like linear combination of $E_{\mathrm{x}}^{\mathrm{HF,SR}}(\omega)$ at
$\omega=0.11$~bohr$^{-1}$ relative to PBE exchange. For
energy constraints, the stoichiometric coefficients combine exchange energies; for
occupation-response constraints, they combine derivatives of those exchange energies with
respect to frontier orbital occupations.

For the ionization-potential and electron-affinity systems, three related constraint families
combine exchange energy and occupation response. To state the actual exchange-only labels,
let $E_{\mathrm{x}}^A$ denote either $E_{\mathrm{x}}^{\mathrm{HF,SR}}(\omega)$ or $E_{\mathrm{x}}^{\mathrm{PBE}}$, and define
$\epsilon_{\mathrm{x},p}^A=\partial E_{\mathrm{x}}^A/\partial f_p$. For the ionization branch, the three combinations represent the exchange contribution to the ionization potential (IP), the residual associated with the generalized Koopmans condition (GKC), and a piecewise quadraticity (PQ) residual:
\begin{align}
&\Delta_{\mathrm{IP},\mathrm{x}}^A
  = E_{\mathrm{x}}^A(N{-}1)-E_{\mathrm{x}}^A(N), \nonumber\\
&\Delta_{\mathrm{GKC},\mathrm{x}}^A
  = \epsilon_{\mathrm{x},\mathrm H}^A(N)-\epsilon_{\mathrm{x},\mathrm L}^A(N{-}1), \nonumber\\
&\Delta_{\mathrm{PQ},\mathrm{x}}^A
  = \Delta_{\mathrm{IP},\mathrm{x}}^A
   + \tfrac12\left[\epsilon_{\mathrm{x},\mathrm H}^A(N)
   + \epsilon_{\mathrm{x},\mathrm L}^A(N{-}1)\right].
\label{eq:si_training_combinations}
\end{align}
The electron-affinity constraints use the analogous electron-addition occupations. In every case,
the supervised target is the difference between the short-range Hartree--Fock and PBE values
of the corresponding combination,
\begin{equation}
y_X = \Delta_{X,\mathrm{x}}^{\mathrm{HF,SR}}-\Delta_{X,\mathrm{x}}^{\mathrm{PBE}},
\qquad X\in\{\mathrm{IP},\mathrm{GKC},\mathrm{PQ}\}.
\label{eq:si_training_residuals}
\end{equation}
Thus the generalized Koopmans constraints are differences of occupation derivatives for the neutral
HOMO and the adjacent charged-state frontier orbital, whereas the piecewise quadraticity constraints
combine an exchange energy change with the two endpoint derivatives. For a quadratic
dependence on the frontier orbital occupation $f\in[0,1]$,
$E_{\mathrm{x}}^A(f)=c_0+c_1f+c_2f^2$ and $\Delta_{\mathrm{GKC},\mathrm{x}}^A=2c_2$. Moreover,
$E_{\mathrm{x}}^A(1)-E_{\mathrm{x}}^A(0)$ equals the average of the two endpoint derivatives, so
$\Delta_{\mathrm{PQ},\mathrm{x}}^A=0$. Exchange energies are further constrained using
W4-11 atomization stoichiometries, reference atoms, and the cohesive energies of the 9 solids.
Separate W4-11 data for frontier orbital occupations constrain the corresponding occupation
derivatives.

Additional uniform gas targets softly constrain the model toward the exact short-range exchange
limit. Noise factors were used to balance energy and occupation-response constraints.

\subsection*{Computational details}

\paragraph{DFT software and RS-CIDER implementation.}
Molecular benchmark calculations used PySCF; periodic calculations of band gaps,
oxides, polarons, and batteries used GPAW (plane-wave basis with PAW datasets) for
RS-CIDER. HSE06 and PBE reference energies for the battery calculations were obtained using VASP (PAW). HSE06 used $\alpha=0.25$ exact short-range exchange with $\omega=0.11~\mathrm{bohr}^{-1}$. Unless otherwise stated, RS-CIDER calculations used the ciderpress implementation with $\mathrm{xmix}=0.25$ (Eq.~\eqref{eq:methods_deploy}) and were performed self-consistently.

\paragraph{GMTKN55 (Fig.~\ref{fig:gmtkn55}).}
The GMTKN55 calculations used PySCF with the \texttt{def2-QZVPPD} basis~\cite{rappoport_property-optimized_2010} and the auxiliary basis \texttt{def2-universal-jkfit} for density fitting~\cite{weigend_hartreefock_2008}. DFT-D4 dispersion~\cite{caldeweyher_generally_2019} with HSE06 parameters
was added consistently to both RS-CIDER and the HSE06 reference. All calculations used an SCF convergence threshold of $10^{-9}$~Ha. The mean absolute deviation in the left panel of Fig.~\ref{fig:gmtkn55} was pooled directly over a reaction-wise held-out
test set of 1113 reactions in 52 subsets. This partition was obtained from the complete 1505-reaction catalogue by excluding the 392 unique GMTKN55 reaction stoichiometries represented among the 1238 supervised training constraints. The split is reaction-wise, so reuse of an individual molecular system does not by itself constitute reaction overlap. Comparison energies for PBE-D4, r$^2$SCAN-D4, PBE0-D4, and $\omega$B97M-V were computed with PySCF using the same orbital basis~\cite{Perdew1996_3865,Furness2020_8208,adamo_toward_1999,mardirossian_omega_2016}. The $\omega$B97M-V data include its VV10 non-local-correlation term;
the other labels identify their corresponding D4-corrected energies.

For the right panel of Fig.~\ref{fig:gmtkn55}, reaction energies for the same 1113 held-out reactions in 52 subsets were calculated from the corresponding system total energies using the GMTKN55 reaction stoichiometries. The reported weighted score was
\begin{equation}
\mathrm{WTMAD\mbox{-}2}_{\mathrm{held\mbox{-}out}}=
\frac{1}{\sum_i N_i}
\sum_i N_i\frac{C_{\mathrm{WTMAD2}}}
{\overline{|\Delta E|}_i}\,\mathrm{MAD}_i,
\end{equation}
where $N_i$ and $\mathrm{MAD}_i$ were evaluated on the common held-out reactions in subset
$i$, while $\overline{|\Delta E|}_i$ was calculated from the full reference-energy definition
of that subset. The normalization constant
$C_{\mathrm{WTMAD2}}=56.84~\mathrm{kcal\,mol^{-1}}$ is the value introduced in the original GMTKN55 definition of WTMAD-2~\cite{goerigk_look_2017}.

\paragraph{Solid-state band gaps (Fig.~\ref{fig:gaps}).}
The RS-CIDER gaps were calculated self-consistently with GPAW through the ciderpress interface in plane-wave mode with $E_\mathrm{cut}=520$~eV and a $\Gamma$-centered even $k$-point mesh with a linear density of 12 points per \AA$^{-1}$. Fermi--Dirac smearing was $0.01$~eV; for Ar, the smearing was set to zero and the number of bands to $200\%$. The GPAW energy-convergence setting was $10^{-6}$~eV per valence electron. The reported generalized Kohn--Sham gaps were
obtained as CBM--VBM eigenvalue differences using \texttt{ase.dft.bandgap}. Structures were ICSD-keyed CIFs from a curated band-gap subset of 17 semiconductors and insulators.

The PBE, r$^2$SCAN, and earlier CIDER values were taken from the GPAW calculations of Bystrom and Kozinsky~\cite{PhysRevB.110.075130}, which used the same plane-wave cutoff and linear $k$-point density. The HSE06 values were taken from the VASP calculations reported in the same work, using a $4000/N_\mathrm{atom}$ $k$-point-mesh criterion; that work reported changes below
$0.01$~eV relative to denser meshes.

\paragraph{Cu--O phase diagram (Fig.~\ref{fig:cuo_phase_diagram}).}
The phase diagram considers Cu (metal), Cu$_2$O, CuO, and O$_2$ as the oxygen reference. CuO was treated in a collinear antiferromagnetic configuration; antiferromagnetic order in CuO is experimentally established~\cite{PhysRevB.38.174}. The four Cu sites were initialized in a
collinear antiferromagnetic pattern with alternating spin directions.
Both functionals used the same Materials Project structures~\cite{jain_commentary_2013}, each relaxing its own cell.
\emph{HSE06:} VASP~6.4.2 (PAW, $\alpha=0.25$, $\omega=0.11$~bohr$^{-1}$,
$E_\mathrm{cut}=800$~eV for all phases); $\Gamma$-centered $k$-meshes $12^3$ (Cu), $8^3$ (Cu$_2$O), $6^3$ (CuO), with O$_2$ a triplet in a $10\times 10\times 10$~\AA{} box. \emph{RS-CIDER:} GPAW through the ciderpress interface, self-consistent single points at plane-wave cutoff $1000$~eV; $\Gamma$-centered $k$-meshes $12^3$, $10^3$, $6^3$ for Cu, Cu$_2$O, CuO. The converged RS-CIDER Cu moment in CuO is $0.63~\mu_B$.

For each condensed phase $i$, the phase diagram compares the grand potential per Cu atom,
\begin{equation}
\Phi_i(T,p_{\mathrm{O}_2})
=
\frac{E_i-N_{\mathrm{Cu},i}\mu_{\mathrm{Cu}}
-N_{\mathrm{O},i}\mu_{\mathrm{O}}(T,p_{\mathrm{O}_2})}
{N_{\mathrm{Cu},i}},
\label{eq:si_cuo_grand_potential}
\end{equation}
with $\mu_{\mathrm{Cu}}=E_{\mathrm{Cu}}^{\mathrm{bulk}}/4$. The oxygen chemical potential is
\begin{equation}
\begin{aligned}
&\mu_{\mathrm{O}}(T,p_{\mathrm{O}_2})
=\frac12E_{\mathrm{O}_2}
+\frac12\left[H_{\mathrm{O}_2}^{\circ}(T)-H_{\mathrm{O}_2}^{\circ}(0)
-T S_{\mathrm{O}_2}^{\circ}(T)\right]\\
&\qquad\qquad\qquad\quad
+\frac12k_{\mathrm B}T\left[
\ln\!\left(\frac{p_{\mathrm{O}_2}}{1~\mathrm{atm}}\right)
+\ln(1.01325)\right],
\end{aligned}
\label{eq:si_oxygen_chemical_potential}
\end{equation}
where the second logarithm converts the plotted pressure coordinate from atm to the $1$-bar standard state used by NIST--JANAF. Standard-state enthalpy increments and entropies were linearly interpolated from the NIST--JANAF
thermochemical data~\cite{https://doi.org/10.18434/t42s31}. The method-specific DFT energy $E_{\mathrm{O}_2}$ was used in Eq.~\eqref{eq:si_oxygen_chemical_potential}, together with the NIST--JANAF thermal contributions and the ideal-gas pressure term; the phase with the lowest
$\Phi_i$ is assigned as stable.

\paragraph{Polaron localization (Fig.~\ref{fig:polaron}).}
\emph{TiO$_2$ oxygen-vacancy electron polaron:} We used a $2\times2\times3$ rutile supercell (72 atoms; 71 in the $V_\mathrm{O}$ cell) and $\Gamma$-point sampling, as in De\'ak \emph{et al.}~\cite{PhysRevB.86.195206}. The starting lattice was taken from the Materials Project rutile structure~\cite{jain_commentary_2013} and held fixed during ionic relaxation ($\mathrm{ISIF}=2$). PBE, PBE+U, and HSE06 used VASP (PAW Ti\_pv and O; $E_\mathrm{cut}=400$~eV for HSE06, $520$~eV for PBE/PBE+U). The PBE+$U$ calculation used the rotationally invariant Dudarev formulation with
$U_{\mathrm{eff}}=U-J=4.2$~eV on Ti $3d$ states~\cite{PhysRevB.57.1505}. RS-CIDER used GPAW with plane-wave $1000$~eV (grid spacing $\sim 0.097$~\AA), $\mathrm{xmix}=0.25$. All calculations used $\mathrm{ISPIN}=2$. The left panel of Figure~\ref{fig:polaron} reports site-resolved absolute local magnetic moments
for the $q=+1$ charge state in a common atom order. All four calculations used the same initial structure and preserved that order. \emph{MgO hole polaron:} conventional 2$\times$2$\times$2 supercell (64 atoms, Mg$_{32}$O$_{32}$),
charge state $q=+1$, $\mathrm{ISPIN}=2$, $\Gamma$-only. HSE06 used VASP~6.4.2 ($E_\mathrm{cut}=520$~eV); RS-CIDER used GPAW with plane-wave $1000$~eV (grid spacing $\sim 0.165$~\AA). For each functional, the exchange fraction $\alpha_k$ was determined non-empirically from the zero crossing of a linear fit to the corrected charged--neutral endpoint eigenvalue difference, the condition associated with a piecewise-linear total energy through Janak's theorem, with Falletta--Wiktor--Pasquarello (FWP) image-charge corrections~\cite{PhysRevB.102.041115}. We used a Gaussian model-charge width of $1.4$~bohr and the PBE MgO dielectric constants $\varepsilon_0=10.73$ and $\varepsilon_\infty=2.77$ reported by Falletta and Pasquarello~\cite{falletta_polarons_2022}. Two lattice protocols were used. In the common-lattice protocol, the same PBE-relaxed cell was held fixed for both functionals. Applying the piecewise-linearity tuning procedure on this cell gave
$\alpha_k=0.4378$ for HSE06 and $0.5740$ for RS-CIDER. In the per-functional protocol, each functional first relaxed the pristine MgO cell using the value of $\alpha_k$ obtained from the common-lattice tuning. The tuning procedure was then repeated on the resulting fixed cell, giving $\alpha_k=0.4527$ for HSE06 and $0.5978$ for RS-CIDER. The polaron formation energy was evaluated with two estimators that coincide for a perfectly piecewise-linear functional~\cite{doi:10.1021/acs.jctc.4c00999}: a $\Delta$SCF total-energy difference and an eigenvalue-based expression. Their difference $|\mathrm{dev}|$ measures the residual non-linearity (Table~\ref{tab:polaron_ef}); it is small for HSE06 ($\le 0.02$~eV) and larger for RS-CIDER ($\approx 0.25$~eV), while the RS-CIDER and HSE06 $\Delta$SCF formation energies agree closely under both lattice protocols, differing by $0.052$~eV for the common-lattice protocol and $0.044$~eV for the per-functional protocol.

For the common-lattice fits in the right panel of Fig.~\ref{fig:polaron}, HSE06 used $\alpha=0.35$ and $0.50$ as the fit endpoints, with $\alpha=0.40$ as an intermediate check. RS-CIDER used
$\alpha=0.40$ and $0.60$ as the fit endpoints, with $\alpha=0.45$ and $0.50$ as intermediate checks. A separate RS-CIDER calculation at the fitted value $\alpha=0.574$ gave $\Delta\varepsilon=+0.024$~eV, consistent with the fitted zero crossing. The complete values are listed in Table~\ref{tab:polaron_alpha_scan}.

\begin{table}[ht]
\centering
\caption{Common-lattice MgO scan values used to select the exchange mixing in
the right panel of Fig.~\ref{fig:polaron}. The corrected polaron-level difference is
$\Delta\varepsilon=[\varepsilon_p(q)+\varepsilon_{\mathrm{cor},qq}]
-[\varepsilon_p(0)+\varepsilon_{\mathrm{cor},0q}]$. The main-text fit uses only the two rows marked as endpoints for each functional.}
\label{tab:polaron_alpha_scan}
\begin{tabular}{l r r l}
\hline
Functional & $\alpha$ & $\Delta\varepsilon$ (eV) & Role \\
\hline
HSE06     & 0.35  & $-0.758$ & fit endpoint \\
HSE06     & 0.40  & $-0.372$ & intermediate check \\
HSE06     & 0.50  & $+0.537$ & fit endpoint \\
RS-CIDER  & 0.40  & $-1.411$ & fit endpoint \\
RS-CIDER  & 0.45  & $-0.996$ & intermediate check \\
RS-CIDER  & 0.50  & $-0.680$ & intermediate check \\
RS-CIDER  & 0.60  & $+0.211$ & fit endpoint \\
RS-CIDER  & 0.574 & $+0.024$ & calculation at fitted $\alpha_k$ \\
\hline
\end{tabular}
\end{table}

\begin{table}[ht]
\centering
\caption{MgO hole-polaron ($q=+1$) formation energy from two estimators that coincide for a perfectly piecewise-linear functional: a $\Delta$SCF total-energy difference,
$E_f(\Delta\mathrm{SCF})$, and an eigenvalue-based expression, $E_f(\mathrm{eig})$, both with FWP finite-size corrections. $|\mathrm{dev}|=|E_f(\Delta\mathrm{SCF})-E_f(\mathrm{eig})|$ gauges
the residual non-linearity. ``Common'' uses the shared PBE-relaxed cell; ``own'' uses each functional's separately relaxed cell and the value of $\alpha_k$ obtained by repeating the tuning procedure on that fixed cell. The HSE06 common-lattice entry is interpolated from the exchange-fraction scan; the remaining entries were calculated directly at their fitted $\alpha_k$ values.}
\label{tab:polaron_ef}
\begin{tabular}{l l r r r r}
\hline
Lattice & Functional & $\alpha_k$ & $E_f(\Delta\mathrm{SCF})$ & $E_f(\mathrm{eig})$ & $|\mathrm{dev}|$ \\
        &            &            & (eV) & (eV) & (eV) \\
\hline
common & HSE06    & 0.4378 & $-0.465$ & $-0.477$ & 0.012 \\
common & RS-CIDER & 0.5740  & $-0.413$ & $-0.653$ & 0.240 \\
own    & HSE06    & 0.4527 & $-0.383$ & $-0.401$ & 0.018 \\
own    & RS-CIDER & 0.5978 & $-0.339$ & $-0.592$ & 0.252 \\
\hline
\end{tabular}
\end{table}

\paragraph{Point-defect formation energy (Fig.~\ref{fig:vacancy}).}
Trigonal Se, $3\times 3\times 3$ supercells: 81 Se in bulk, 80 in the neutral vacancy cell, and 82 in the neutral interstitial cell. The general RS-CIDER model used GPAW (plane-wave, $E_\mathrm{cut}=600$~eV, no SOC); PBE and HSE06 used VASP ($E_\mathrm{cut}=300$~eV; HSE06 at 25\% short-range exact exchange, $\omega=0.11$~bohr$^{-1}$). All calculations included D3 dispersion
with the parameters of the HSE06 reference ($s_8=0.928$, $r_{s6}=1.287$) and a $\Gamma$-centered $2\times 2\times 2$ $k$-mesh; ionic relaxations used VASP $\mathrm{IBRION}=2$ / GPAW BFGS.
The published HSE06+SOC structures and formation energies of Kavanagh \emph{et al.} define the reference geometry and benchmark values~\cite{10.1039/d4ee04647a}. The Se-rich formation energies are $E_f(V)=E(\mathrm{vac},80)-\tfrac{80}{81}\,E(\mathrm{bulk},81)$ and $E_f(\mathrm{Int})=E(\mathrm{int},82)-\tfrac{82}{81}\,E(\mathrm{bulk},81)$, with defect and bulk taken on the same lattice.

For $q=0$ the formation energy needs no charge correction and no cross-code valence-band reference, so the GPAW (RS-CIDER) and VASP (PBE, HSE06) values are directly comparable. Two
controls minimize additional methodological differences. (i)~All four methods use identical D3 parameters, providing a controlled dispersion treatment. (ii)~Computing HSE06 with and without spin--orbit coupling (otherwise identical) changes the formation energies by only $0.019$~eV ($V_{\mathrm{Se}}$) and $<0.001$~eV (Int), and the SOC- and no-SOC-relaxed lattices differ by $<0.06\%$.

We report three geometry protocols (Table~\ref{tab:vacancy}): (i)~a single point with all atoms at the HSE06 geometry, probing electronic-energy differences without geometry-relaxation differences; (ii)~ionic relaxation ($\mathrm{ISIF}=2$) at the fixed reference cell; and (iii)~ionic relaxation at each functional's independently relaxed bulk lattice. For the RS-CIDER vacancy in protocol (iii), a clean BFGS calculation converged to $f_{\max}=0.044$~eV/\AA{} before a tighter single-point calculation gave $E_f=1.466$~eV.
At fixed cell the functionals relax to nearly the same ionic geometry---the RS-CIDER and PBE structures lie within $0.07$~\AA{} (RMSD) of the HSE06 geometry for both defects (RS-CIDER $0.04$/$0.02$, PBE $0.05$/$0.07$~\AA{} for vacancy/interstitial), so these small geometric differences alone cannot explain the energy spread. Consistently, the residual forces at the HSE06 geometry are largest for PBE ($\sim 0.7$~eV/\AA), smaller for RS-CIDER ($\sim 0.4$), and near zero for HSE06. Under protocol (iii) PBE's relaxed lattice over-binds (its $a$ axis contracts $4.4\%$), shifting its bulk reference by $1.6$~eV and producing the large protocol-(iii) swings, whereas RS-CIDER's lattice expands only $1.5\%$ and its formation energies stay near HSE06. The one case where PBE is fortuitously close is the interstitial at the fixed HSE06 geometry (Table~\ref{tab:vacancy}); on relaxation PBE drives that energy negative while RS-CIDER converges toward HSE06.

\begin{table}[ht]
\centering
\caption{Formation energies $E_f$ (eV) of the neutral Se vacancy and interstitial in trigonal Se ($q=0$) under three geometry protocols: all atoms at the HSE06 geometry (``fixed''), ions relaxed in the fixed reference cell (``reference lattice''), and ions relaxed at each functional's own optimized bulk lattice (``own bulk''). The HSE06 row repeats the published HSE06+SOC reference values of Kavanagh \emph{et al.} across the three protocol columns~\cite{10.1039/d4ee04647a}.}
\label{tab:vacancy}
\begin{tabular}{l l r r r}
\hline
Defect & Functional & fixed & reference lattice & own bulk \\
\hline
$V_{\mathrm{Se}}^{0}$  & PBE      & 1.080 & 0.176 & 1.274 \\
                       & RS-CIDER & 1.362 & 1.118 & 1.466 \\
                       & HSE06    & 1.525 & 1.525 & 1.525 \\
\hline
$\mathrm{Int}_{\mathrm{Se}}^{0}$ & PBE & 0.927 & $-0.107$ & 1.510 \\
                       & RS-CIDER & 1.140 & 0.949 & 0.949 \\
                       & HSE06    & 0.823 & 0.823 & 0.823 \\
\hline
\end{tabular}
\end{table}

\paragraph{Battery intercalation voltages (Fig.~\ref{fig:voltage}).}
The four olivine cells comprise the lithiated forms LiFePO$_4$ and LiMnPO$_4$ (28 atoms each) and the delithiated forms FePO$_4$ and MnPO$_4$ (24 atoms each); bulk bcc Li (one atom) served as the lithium reference. The olivine geometries were the DFT+$U$+$V$-relaxed structures reported by Timrov \emph{et al.}~\cite{PRXEnergy.1.033003}, which were also used for the HSE06 calculations in that study. Antiferromagnetic AF1 ordering was imposed on the transition-metal sites in the pattern $(+,+,-,-)$. PBE and HSE06 reference energies used VASP~6.4.2 (PAW Fe\_pv / Mn\_pv / O / P / Li\_sv) with $E_\mathrm{cut}=1100$~eV for PBE and $E_\mathrm{cut}=600$~eV for HSE06. RS-CIDER energies used GPAW + ciderpress with $\mathrm{xmix}=0.25$, self-consistent SCF, and a uniform $E_\mathrm{cut}=1100$~eV. $k$-meshes were $6\times 8\times 10$ $\Gamma$-centered for the olivines and $10\times 10\times 10$ for bulk Li. Voltages were computed as $V = -[E(\mathrm{LiMPO}_4) - E(\mathrm{MPO}_4) - 4\,E(\mathrm{Li})]/(4e)$ with $n=4$ Li per unit cell and $e$ the elementary positive charge.

Mn charge transfer was evaluated by applying the grid-based Bader partitioning algorithm~\cite{HENKELMAN2006354} to the converged charge densities of LiMnPO$_4$ and MnPO$_4$. The net Mn Bader charge was averaged over the four Mn sites in each cell, and the reported Mn-centered redox transfer was defined as $\Delta q_{\mathrm{Mn}}=\overline q_{\mathrm{Mn}}(\mathrm{MnPO}_4) -\overline q_{\mathrm{Mn}}(\mathrm{LiMnPO}_4)$; positive $\Delta q_{\mathrm{Mn}}$ therefore denotes a loss of Mn-centered electron density upon delithiation.

\paragraph{Computational cost (Fig.~\ref{fig:cost}).}
Diamond C: 8-atom perfect cubic cell ($a=3.567$~\AA) and a single-vacancy 7-atom variant; GPAW plane-wave mode with a 500~eV cutoff (grid spacing $0.1784$~\AA); $4\times 4\times 4$
$\Gamma$-centered Monkhorst--Pack $k$-mesh; Fermi--Dirac smearing $0.01$~eV; and energy convergence threshold $10^{-5}$~eV per valence electron. Plain PBE and RS-CIDER were each run three times on the same physical node using 32 MPI ranks. The HSE06 timings were taken from existing GPAW HybridXC calculations with identical numerical settings and the same node type, using $\alpha=0.25$ and $\omega=0.11$~bohr$^{-1}$.

Per-SCF-step wall times are the GPAW Timing report's SCF-cycle total divided by the number of completed iterations.

\begin{table}[ht]
\centering
\caption{Per-SCF-step wall times (s) for the 8-atom diamond-C cell and the 7-atom C-vacancy cell, with 32 MPI ranks used for each calculation. PBE and RS-CIDER show three paired repeats, with medians in bold parentheses. The HSE06 values are averages over completed SCF steps.}
\label{tab:cost_timings}
\footnotesize
\setlength{\tabcolsep}{4pt}
\begin{tabular}{lcc}
\hline
Method & Perfect cell & C vacancy \\
\hline
PBE & 0.0609, 0.0602, 0.0593 (\textbf{0.0602})
    & 0.0567, 0.0567, 0.0562 (\textbf{0.0567}) \\
RS-CIDER & 0.3300, 0.3303, 0.3308 (\textbf{0.3303})
         & 0.3324, 0.3344, 0.3398 (\textbf{0.3344}) \\
HSE06 & 58.9781 & 47.9680 \\
\hline
\end{tabular}
\end{table}

\subsection*{Solid-state band-gap data}

Table~\ref{tab:solid_gaps} reports the calculated and experimental band gaps of the 17 semiconductors and insulators used in the main-text comparison (Fig.~\ref{fig:gaps}). 

The RS-CIDER gaps were obtained from self-consistent GPAW calculations using a 520~eV plane-wave cutoff and the smallest even, $\Gamma$-centered $k$-point mesh with a linear density of at least 12 $k$ points per \AA$^{-1}$. Each band gap was evaluated as the difference between the conduction-band minimum and valence-band maximum on this $k$-point mesh.
The PBE, r$^2$SCAN, earlier CIDER, and HSE06 band gaps were taken from Bystrom and Kozinsky~\cite{PhysRevB.110.075130}. In that work, the PBE, r$^2$SCAN, and CIDER calculations used the same cutoff and $k$-point-density criterion, whereas the HSE06 calculations used VASP, with the $k$-point mesh set to contain 4000 $k$ points divided by the number of atoms in the unit cell.

Across this set, RS-CIDER reproduces the HSE06 reference gaps with a mean absolute deviation of $0.197$~eV (mean signed deviation $-0.102$~eV), roughly halving the $0.362$~eV mean absolute deviation of the earlier CIDER functional, which targets PBE0. HSE06 is the reference for RS-CIDER; HSE06 itself underestimates the experimental gaps of the widest-gap systems (e.g., CaO, NaCl, LiF, and Ar), so the experimental column is provided only for context.

\begin{table}[ht]
\centering
\caption{Calculated and experimental band gaps (in eV) for the solid-state benchmark set. PBE, r$^2$SCAN, CIDER, RS-CIDER, and HSE06 are computed values. Experimental gaps are shown for context, whereas HSE06 is the reference for assessing RS-CIDER fidelity. RS-CIDER values were calculated in this work with GPAW; the remaining calculated values are from Bystrom and Kozinsky~\cite{PhysRevB.110.075130}, and experimental values are from Borlido \emph{et al.}~\cite{borlido_large-scale_2019}.}
\label{tab:solid_gaps}
\begin{tabular}{l r r r r r r}
\hline
Material & PBE & r$^2$SCAN & CIDER & RS-CIDER & HSE06 & Exp. \\
\hline
Si   &  0.57 &  0.76 &  1.05 &  1.14 &  1.15 &  1.17 \\
Ge   &  0.12 &  0.48 &  0.71 &  0.97 &  0.81 &  0.74 \\
InP  &  0.69 &  1.12 &  1.47 &  1.67 &  1.52 &  1.42 \\
GaAs &  0.59 &  1.08 &  1.36 &  1.68 &  1.43 &  1.52 \\
CdSe &  0.71 &  1.19 &  1.48 &  1.68 &  1.69 &  1.74 \\
BP   &  1.25 &  1.42 &  1.71 &  1.84 &  1.98 &  2.10 \\
GaP  &  1.61 &  1.88 &  2.09 &  2.30 &  2.29 &  2.35 \\
CdS  &  1.20 &  1.65 &  2.02 &  2.22 &  2.27 &  2.48 \\
GaN  &  1.88 &  2.33 &  2.65 &  2.76 &  3.18 &  3.50 \\
ZnS  &  2.13 &  2.69 &  3.04 &  3.30 &  3.37 &  3.72 \\
C    &  4.13 &  4.34 &  4.70 &  4.84 &  5.33 &  5.50 \\
BN   &  4.21 &  4.79 &  5.08 &  5.38 &  5.63 &  5.96 \\
CaO  &  3.68 &  4.23 &  4.53 &  5.05 &  5.32 &  6.88 \\
MgO  &  4.73 &  5.63 &  6.11 &  6.22 &  6.42 &  7.67 \\
NaCl &  5.11 &  5.90 &  6.32 &  6.54 &  6.45 &  8.75 \\
LiF  &  9.07 & 10.05 & 10.40 & 10.76 & 11.39 & 13.60 \\
Ar   &  8.71 &  9.61 &  9.71 & 10.51 & 10.36 & 14.15 \\
\hline
\end{tabular}
\end{table}

\begingroup
\catcode`\&=12
\bibliographystyle{unsrtnat-preserve-title}
\bibliography{sample}
\endgroup

\end{document}